\documentclass[aps,prl,superscriptaddress,twocolumn]{revtex4-2}

\usepackage{amsmath}
\usepackage{amssymb}
\usepackage{graphicx}
\usepackage{color}
\usepackage{bm}
\usepackage{subfigure}
\usepackage{multirow}
\usepackage{latexsym}
\usepackage{mathptmx}
\usepackage{comment}

\begin{document}
\title{Enhanced phoretic self-propulsion of active colloids through surface charge asymmetry}

\author{Ahis Shrestha}

\affiliation{Center for Computation and Theory of Soft Materials, Northwestern University, Evanston, IL 60208}

\affiliation{Department of Physics and Astronomy, Northwestern University, Evanston, IL 60208}

\author{Monica Olvera de la Cruz}

\affiliation{Center for Computation and Theory of Soft Materials, Northwestern University, Evanston, IL 60208}
\affiliation{Department of Material Science and Engineering, Northwestern University, Evanston, IL 60208}
\affiliation{Department of Physics and Astronomy, Northwestern University, Evanston, IL 60208}


\begin{abstract}
	
Charged colloidal particles propel themselves through asymmetric fluxes of chemically generated ions on their surface. We show that asymmetry in the surface charge distribution introduces a new mode of self-phoretic motion for chemically active particles that produce ionic species. Particles of sizes smaller than or comparable to the Debye length achieve directed self-propulsion through surface charge asymmetry even when ionic flux is uniform over the particle surface. Janus nanoparticles endowed with both surface charge and ionic flux asymmetries results in enhanced propulsion speeds of the order of $\mu$m/s or higher. Our work provides a theoretical framework to quantitatively determine the velocity of asymmetrically charged nanoparticles undergoing ionic self-diffusiophoresis, and suggests an avenue for specifying surface properties that optimize and regulate self-propulsion in ionic media.

\end{abstract}

\maketitle

Many microorganisms exhibit locomotion in viscous fluid environments typically driven by asymmetric surface activity \cite{Bray2000, *Lauga2009}. Inspired by such biological systems, micro- and nanomotors have been developed with a wide range of functionalities for targeted cargo transport and various other biomedical or environmental applications \cite{Moran2017,Gao2014, *Soler2014}. Directed motion of colloidal micro- or nanoparticles has been achieved through externally imposed fields or gradients as in electrophoresis or diffusiophoresis \cite{Anderson1989,Velegol2016}. Alternatively, active colloidal particles, namely Janus micro- or nanomotors, self-propel by harnessing locally stored chemical energy \cite{Marchetti2013,Illien2017}. These chemically active motors can be designed with specified surface features to enhance self-propulsion \cite{Golestanian2005, *Golestanian2007, Zhang2017}. 

Phoretic active colloids propel themselves through a self-generated concentration gradient of solute, a process known as self-diffusiophoresis. Local chemical gradients are generally mediated by surface reactions that induce a non-uniform flux of chemical species, which allow for phoretic movement resulting from the chemical transport in the surrounding fluid \cite{Moran2017,Illien2017}. Such self-propelling particles may also have a certain intrinsic charge, in concurrent, with a flux of charged ionic species being released or absorbed at the surface. Moreover, the bulk solution itself may have some background ionic concentration. In this case, an electrostatic contribution arises from the interaction between the charged surface and the mobile ions in the fluid. This couples with the ionic transport and the hydrodynamics, leading to a different type phoretic self-propulsion called ionic self-diffusiophoresis \cite{Illien2017, Zhou2018,DeCorato2020,Asmolov2022, Brown2017}. 

\begin{figure}[b]
	\center
	\includegraphics[width=0.8\columnwidth]{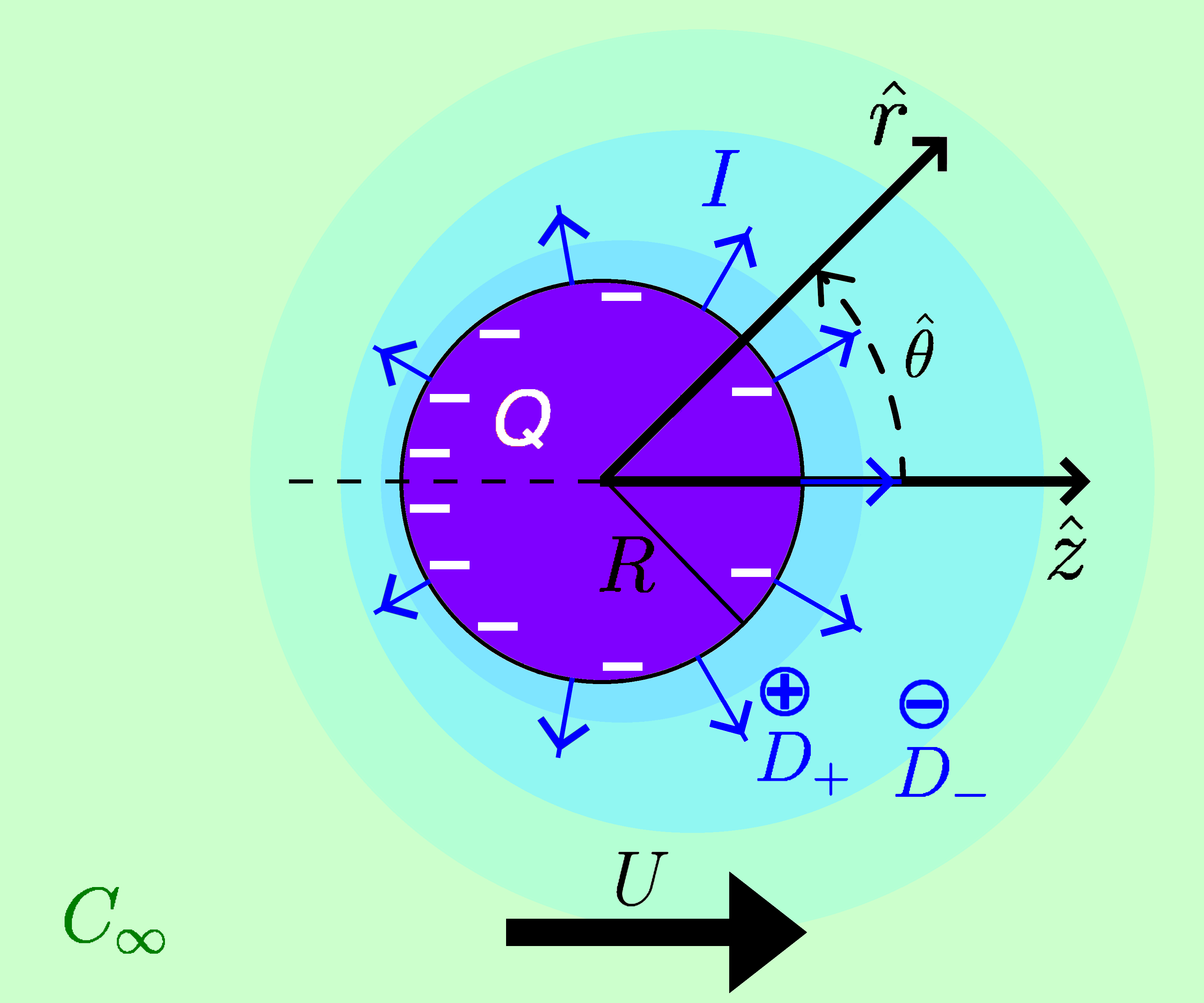} 
	\caption{Schematic of ionic self-diffusiophoresis. A spherical colloidal particle of radius $R$ and net negative charge $Q$ that releases positive and negative ion species at a net rate $I$ from the surface with respective diffusivities $D_{+}$ and $D_{-}$. The colloid has asymmetric distributions of surface charge and ionic flux, and drifts with velocity $U$ in the z-direction inside a solution with bulk ionic concentration $C_\infty$. 
	}
	\label{f1}
\end{figure}

In recent years, phoretic self-propulsion of charged colloidal particles, driven by the release of ions from their surface, has been realized in the realm of micro- and nanomotors, such as urease-powered colloids or photo-activated AgCl Janus particles \cite{DeCorato2020,Zhou2018, Zhang2017, Hermanova2020, *Safdar2018}. We demonstrate here that a particle undergoing ionic self-diffusiophoresis (see Fig.~\ref{f1}) attains directed self-propulsion uniquely mediated by an asymmetric distribution of surface charge, even when the chemically generated ionic flux is uniform over the surface. This occurs in regimes where the particle size is smaller than or comparable to the Debye length. In such domains, the near-field electrostatic and hydrodynamic contributions resulting from the asymmetry in surface charge become crucial. Moreover, we find nanosized particles bearing both surface charge and ionic flux asymmetries achieve enhanced self-propulsion with optimal speeds that persist at higher ionic concentrations, in contrast to microsized organisms or particles for which such phoretic motion subside \cite{Moran2011}. We hypothesize that this particular type of phoretic self-propulsion mechanism will play a role in the transport of cellular subunits such as bacterial microcompartments (BMC), which are typically 40 to 200 nm in size and have highly heterogeneous shell surfaces \cite{Sutter2021, Kerfeld2005}. BMCs facilitate local chemical reactions that produce ionic fluxes at specific sites as well as carry a non-uniform distribution of charge on their shell \cite{Kerfeld2005, Sommer2017, *Sutter2019}. The ionic environment in cell media may allow for directed self-propulsion of such nanosized components equipped with active surface properties. A quantitative understanding of the physical principles underlying subcellular transport and the significance of electrostatic contributions in this regime that have not been addressed previously are accounted here \cite{Santiago2018,*Lee2014, Ramm2021, *Sear2019}. 

We take a continuum approach \cite{Moran2011,Yariv2011,Brown2017, DeCorato2020, Asmolov2022} to characterize the self-phoretic motion arising form the coupling of electrostatics with ionic transport and fluid flow, and determine the drift velocity of a colloidal particle. We consider a spherical colloid of radius $R$ with a net charge $Q$ suspended in a solution with total background (bulk) ion concentration $C_\infty$. Through chemical reactions on the surface \cite{Zhou2018,DeCorato2020}, the colloid produces different ion species that are released at a total rate $I$ from the surface. We assume the colloid is at a steady state drifting with constant phoretic velocity $U$ in the z-direction (see Fig.~\ref{f1}). The electrostatic interaction between the charged surface of the colloid and the mobile ions, that are being produced at the surface as well as the existing background ions in the solution, is accounted by the Poisson equation 
\begin{equation} \label{eqP} 
    \nabla^2 \psi = - \frac{e}{\varepsilon} \sum_i Z_i C_i  \,,
\end{equation}
where $\psi$ is the electrostatic potential, $e$ is the elementary charge, $\varepsilon$ is the permittivity in water, and $C_i$ is the concentration of the ionic species $i$ with ion valency $Z_i$. Due to the motion of charged ion species in the resulting electric field, an additional electrochemical coupling accompanies the chemical diffusion as prescribed by the Nerst-Planck equations of ionic transport, 
\begin{equation} \label{eqNP} 
    \Vec{\nabla} \cdot \Vec{J_i}= \Vec{\nabla} \cdot \left[ -D_i\Vec{\nabla} C_i - \frac{e Z_i D_i}{k_BT} C_i \Vec{\nabla} \psi  \right] =0  \,,
\end{equation}
where $J_i$ is the flux and $D_i$ is the diffusivity of ion species $i$. We have excluded the advective term in Eq.~(\ref{eqNP}) since the P\'eclet number is low based on the small size and speed of the colloid relative to the diffusivity of the ions. These ranges of colloidal size and speed falls under the low Reynolds number regime as well. Henceforth, the flow of the solution, considered as an incompressible fluid, is described using the Stokes equations
\begin{equation} \label{eqS1}
    \Vec{\nabla} p -\eta \nabla^2 \Vec{v} = \varepsilon \nabla^2 \psi \Vec{\nabla} \psi \,,  \ \ \ \ \ \ \ \Vec{\nabla} \cdot \Vec{v}=0 \,,
\end{equation}
and
where $\Vec{v}$ is the fluid velocity with $p$ the pressure and $\eta$ the viscosity of water. The electrostatic volume force density comes into Eq.~(\ref{eqS1}) as an applied body force in the fluid maintaining momentum balance \cite{DeCorato2020, Asmolov2022}.

Given a spherical particle moving at a constant velocity, we can apply the generalized Lorentz reciprocal theorem for Stokes flow with non-zero body force \cite{Happel1983, *Stone1996, Masoud2019} to directly calculate the phoretic velocity. Following previous works \cite{DeCorato2020,Asmolov2022},  we find that the phoretic velocity is given by   
\begin{equation} \label{eqU}
  U  = \frac{\varepsilon}{6 \pi \eta R} \int_V  \nabla^2 \psi \Vec{\nabla} \psi  \cdot (\hat{v}- \hat{z})\ dV \,,
\end{equation}
where $V$ is the entire fluid volume outside the colloid, and 
\begin{equation}
    \hat{v} =  \left( \frac{3R}{2r} -\frac{R^3}{2r^3}\right) \cos{\theta}  \hat{r} - \left( \frac{3R}{4r} +\frac{R^3}{4r^3} \right) \sin{\theta} \hat{\theta} \,,
\end{equation}
is the solution of the Stokes equations with zero body force for a sphere drifting with unit velocity in the positive z-direction \cite{Landau1987}. This formalism, through which electrostatics, ion transport and fluid flow are coupled, allows us to compute the velocity of a self-phoretic colloidal particle for specified surface and bulk conditions.

We introduce the asymmetry in the surface charge and ion release by defining charge density $\sigma(\theta)$ and ionic flux $j(\theta)$ at the surface that have a polar angle $\theta$ dependence with azimuthal symmetry (see Fig.~\ref{f1}). We consider the background ions in our solution to be a 1:1 binary electrolyte, namely Na$^+$ and Cl$^-$ with respective concentrations $C_{b+}$ and $C_{b-}$. Assuming electroneutrality at the bulk with $\psi \rightarrow 0$ and having no external flux of these background ions, implies a Boltzmann distribution $C_{b \pm}= C_\infty e^{\mp e \psi/ k_B T}/2$ at equilibrium. While on the surface of the colloid, we consider additional ions are generated releasing two monovalent ion species, one positive with $C_+$ and one negative with $C_-$, as in the case of urease-powered motors where ions NH$_{4}^{+}$ and OH$^-$ are being produced \cite{DeCorato2020}. We assume both these ions have the same flux $j(\theta)$ at the surface with the same rate of production such that $I_{\pm}=I/2$ and they diffuse in water with diffusivity $D_\pm$ as $C_{\pm} \rightarrow 0$ at the bulk (see Fig.~\ref{f1}). Furthermore, we have no-slip conditions at the surface-solution interface and the fluid is stagnant at the bulk. More specifically, the set of coupled Poisson-Nerst-Planck-Stokes equations Eq.(\ref{eqP})-(\ref{eqS1}) are subject to the following boundary conditions: (A) $- \varepsilon \Vec{\nabla} \psi \cdot \hat{r} = \sigma(\theta)$, $\Vec{J}_{\pm} \cdot \hat{r} = j(\theta)$, and $\Vec{v}=U \hat{z}$ at the colloid's surface $r=R$, and (B) $\psi=0$, $C_{\pm}=0$ and $\Vec{v}=0$ at the bulk $r \rightarrow \infty$. 

To obtain an analytic approximation for the phoretic velocity, we employ a near-equilibrium perturbation expansion in the Debye-H\"{u}ckel (DH) limit \cite{DeCorato2020}. In particular, we consider regimes of small ionic flux and low surface charge such that $|I|< 2 \pi R (D_++D_-) C_\infty$ and $|Q|<4 \pi R \varepsilon k_BT/e$. This results in Eq.~(\ref{eqP}) being written as the modified DH equation,
\begin{equation} \label{eqDH} 
    \nabla^2 \psi \approx \frac{e}{\varepsilon}(C_{-}- C_{+}) + \kappa^2 \psi \,,
\end{equation}
where $\kappa= \sqrt{e^2 C_\infty / \varepsilon k_B T}$. We estimate $U$ by combining Eq.~(5) with Eq.~(\ref{eqDH}) and solving for $C_\pm$ and $\psi$ from Eqs.~(\ref{eqNP}) and~(\ref{eqDH}) under the boundary conditions (A) and (B) in the leading order of $I$ and $Q$. Based on this continuum model, we quantitatively study the phoretic velocity for varied cases of surface asymmetries and bulk ionic strengths.

\begin{figure*}[t]
	\center
	\includegraphics[width=2\columnwidth]{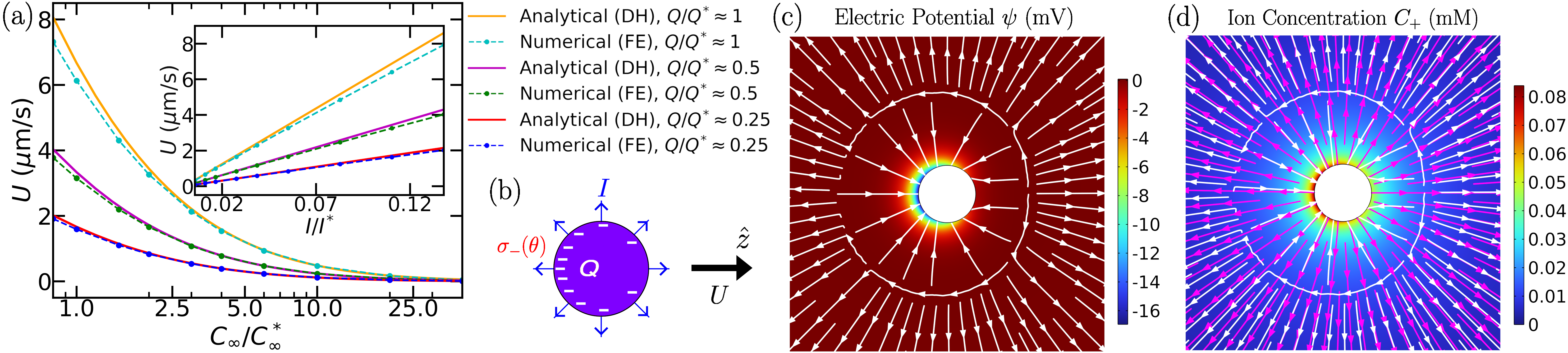} 
	\caption{Self-phoretic motion induced by surface charge asymmetry. (a) The analytical and numerical phoretic speed $U$ as a function of $C_\infty/C_\infty^*$ for $I/I^* \approx 0.05$, and as a function of $I/I^*$ for $C_\infty/C_\infty^* \approx 2$ (inset), both at the indicated values of $Q/Q^*$. We use here the leading order near-equilibrium Debye-H\"{u}ckel (DH) analytical values given by Eq.~(\ref{eqUs}) and the numerical result are from finite element (FE) simulations done in {\it COMSOL Multiphysics} \cite{comsol}. (b) Schematic of the colloid considered here, with uniform ion flux and asymmetric surface charge density $\sigma_-(\theta)=Q(1- \cos{\theta})/4 \pi R^2$, moving in the $\hat{z}$ direction. The color maps with field lines, from the FE simulations in {\it COMSOL}, of (c) the electric potential with electric field and (d) the ion concentration of the positive ion being released $C_+$ with diffusive (magenta) and electrochemical (white) flux, both for $I \approx 0.05 \ I^*$, $Q \approx Q^*$, and $C_\infty \approx C_\infty^*$. In all panels, we use $C_\infty^*=1$ mM, $Q^*= - 4 \pi R \varepsilon k_BT/e$, and $I^* = 2 \pi R (D_++D_-) C^*_\infty$ with $R=20$ nm, $D_+=1.98 \times 10^{-9}$ m$^2/$s, and $D_-=5.27 \times 10^{-9}$ m$^2/$s.        
	}
	\label{f2}
\end{figure*}

We first examine the scenario in which a particle releases ions uniformly from the surface, but has a non-uniform surface charge. In particular, we have a constant flux of $j(\theta)=j_0$ where $j_0=I/8 \pi R^2$ for both the positive and negative ion species. Whereas, we assume a surface charge distribution in which the charges are more dense on one end of the sphere and then gets gradually less dense across the surface towards the opposite end. We consider two configurations: one $\sigma_{+}$ where the charges are more dense on the front ($+\hat{z}$) end, and the other $\sigma_{-}$ where charges are more dense on the back ($-\hat{z}$) end. Namely, we defined the two charge densities as $\sigma_{\pm}(\theta)= \sigma_0 (1 \pm \cos{\theta})$ with $\sigma_0=Q/4 \pi R^2$ to capture this type of surface charge asymmetry. In this case, we get the leading order phoretic velocity for charge densities $\sigma_\pm$ as   
\begin{equation} \label{eqUs}
    U_{\sigma \pm}  = \mp \ I Q \frac{\alpha}{R}  \ f(\kappa R) + \mathcal{O} \left( I^2,Q^2 \right) \,,
\end{equation}
with $\alpha=e(D_+-D_-)/16 \pi \varepsilon \eta D_+ D_- $ and
\begin{equation}\label{eqF}
    f(x)= \frac{x^2(x^2-12)e^x Ei(-x)+(x+3)(x^2-4x+2)}{144(x^2+2x+2)} \,,
\end{equation}
where $Ei$ is the exponential integral function. We observe this leading order velocity that is linear in $I$ and $Q$ vanishes if the positive and negative ions have relatively similar diffusivity as $\alpha \rightarrow 0$. Additionally, we see from Eq.~(\ref{eqF}) that when the size of the particle is small compared to the Debye length $1/\kappa$, it acquires larger speeds which increases as bulk ion concentration is lowered, and as $\kappa R \rightarrow 0$, $f(\kappa R)$ approaches a constant value such that $U_{\sigma, \text{max}} \approx |I Q \alpha| / 48 R$ is the maximum speed in this DH limit. Conversely, if the particle size is much larger than the Debye length as $\kappa R \rightarrow \infty$, $f(\kappa R) \rightarrow 0$ implying a very small vanishing speed.     

We show in Fig.~\ref{f2}(a) the analytical DH approximation of the phoretic speed given by Eq.~(\ref{eqUs}) and numerical results for our model performed using finite element (FE) simulations in {\it COMSOL Multiphysics} software \cite{comsol}. We consider the domain of small ion flux and low surface charge where our analytic approximations are valid, in particular, below the critical limit of net charge $|Q^*| \approx 4 \pi R \varepsilon k_BT/e$ and ion release rate $|I^*| \approx 2 \pi R (D_++D_-) C^*_\infty$ where we set $R=20$ nm and $C^*_\infty = 1$ mM. In this domain, we find in agreement with the FE results that the speed scales linearly with $I$ and $Q$, and promptly decreases with increase in bulk ion concentration $C_\infty$ for a fixed $R$ as predicted by Eq.~(\ref{eqUs}). We also find $U>0$ for a particle of net negative charge $Q<0$ with surface charge density $\sigma_-$ that is releasing ions with distinct diffusivity such that $D_+<D_-$, using Eq.~(\ref{eqUs}) in confirmation with FE simulations. This implies that in this case the particle moves in the positive $z$ direction, away from it's densely charged side (see Fig.~\ref{f2}(b)). Furthermore, for such parameters used in Fig.~\ref{f2} (a), phoretic speeds of $U \gtrsim 1 \ \mu$m/s can be predicted within the DH regime for $\kappa R \lesssim 1$ where the effects of surface charge asymmetry become more prominent.      

To gain insight on the physical means through which the phoretic self-propulsion is being induced here, we need to account for near-field effects \cite{Zhou2018}, specially, when the Debye length is comparable to colloid size. Otherwise, if we only consider the far-field perspective, the system appears to have no asymmetries to cause directed movement as the ion flux is uniform and, consequently, the resulting electric field is symmetric around the particle in this domain beyond the Debye region. Hence, in this far-field limit, one would conclude that the particle will have vanishing velocity insufficient to self-propel. On the contrary, we find here that nanosized particles, at bulk ion concentration such that the Debye length becomes proportionate with its size, can move with significant self-propulsion speeds in the order of $\mu$m/s or higher depending on the amount of ion flux and surface charge. The basis of this motion can be understood by looking at the near-field electrostatics that couples to the hydrodynamics. We find here that the non-uniform surface charge density produces asymmetric electrostatic potential and electric field as shown in Fig.~\ref{f2}(c). This constitutes an asymmetric electrochemical flux in addition to the uniform diffusive flux, and thus the competing net effect results in an non-uniform ionic distribution around the particle, for instance, shown in Fig.~\ref{f2}(d) for the positive ions. This induced ionic gradient gives rise to diffusiophoretic fluid flow near the particle leading to self-propulsion \cite{Anderson1989, Moran2017, Velegol2016}. In contrast with propulsion generated only by asymmetric ionic flux \cite{DeCorato2020, Asmolov2022}, this asymmetry in surface charge offers an alternative and additional mechanism through which a local concentration gradient is created to mediate self-phoretic motion.

\begin{figure}[t!]
	\center
	\includegraphics[width=0.82\columnwidth]{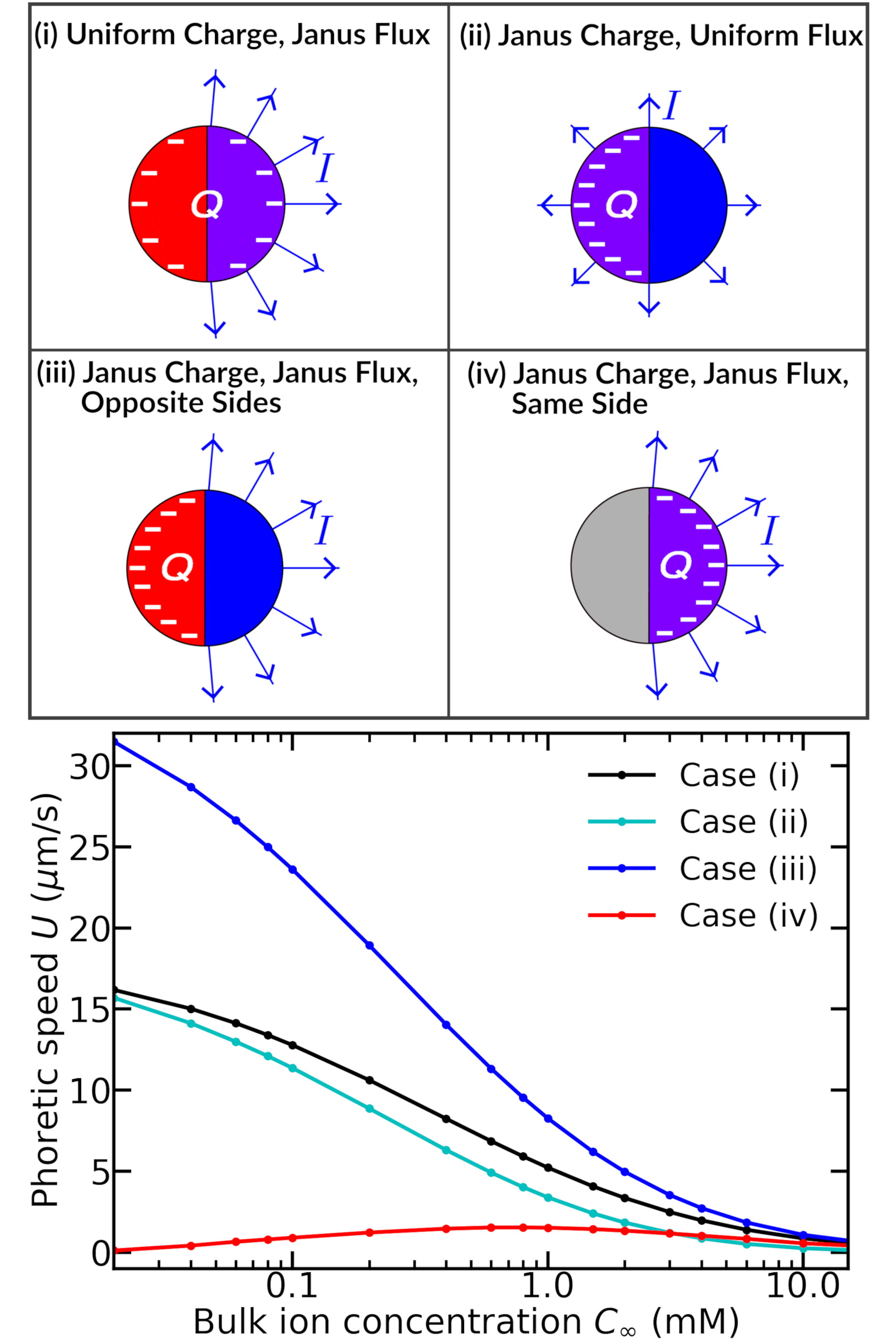} 
	\caption{Enhanced self-propulsion of Janus nanoparticles. Schematic of Janus surface charge and ionic flux configurations (i)-(iv) as describe in text, all with net rate of ions release $I$ and net negative charge $Q$ (top panel). The corresponding phoretic speeds $U$, computed with finite element simulations in {\it COMSOL Multiphysics} \cite{comsol}, for the cases (i)-(iv) as a function of bulk ion concentration $C_\infty$ (bottom panel).  We use here $I \approx 6 \times 10^6$/s and $Q \approx - 60 e$ with $R=20$ nm, $D_+=1.98 \times 10^{-9}$ m$^2/$s, and $D_-=5.27 \times 10^{-9}$ m$^2/$s. 
	}
	\label{f3}
\end{figure}

We next explore the effect of combing surface charge asymmetry with ionic flux asymmetry on the phoretic velocity. Firstly, we employ FE simulations to numerically compute the phoretic velocity of Janus colloids for varying surface charge and ion flux configurations. In particular, we examine four distinct cases of Janus surface charge and ionic flux asymmetries as shown in Fig.~\ref{f3}: (i) ``Uniform Charge, Janus Flux" where the surface charge density is constant on the whole sphere while the ionic flux is finite on one hemisphere and zero on the other, (ii) ``Janus Charge, Uniform Flux" where instead the ionic flux is constant on the whole sphere while the surface charge density is finite on one hemisphere and zero on the other, (iii) ``Janus Charge, Janus Flux, Opposite Sides" where on one hemisphere the surface charge density is finite but ionic flux is zero and vice versa on the other, and (iv) ``Janus Charge, Janus Flux, Same Side" where the surface charge density and ionic flux are both finite on same hemisphere and both zero on the other. We set cases (i)-(iv) such that they all have same $I$ and $Q$. More specifically, we have the uniform surface charge and ion flux $\sigma(\theta)=\sigma_0$ for case (i) and $j(\theta)=j_0$ for case (ii), respectively. The Janus ion flux for cases (i), (iii) and (iv) is $j(\theta)=2j_0$ for $\theta \leq \pi/2$ and $j(\theta)=0$ for $\theta > \pi/2$. While Janus surface charge for cases (ii) and (iii) is $\sigma(\theta)=2\sigma_0$ for $\theta \leq \pi/2$ and $\sigma(\theta)=0$ for $\theta > \pi/2$, and vice versa for case (iv).

We compare and contrast the corresponding phoretic velocities of cases (iii) and (iv) with respect to case (i) and (ii) at different bulk ion concentrations to examine the quantitative differences in phoretic speed resulting from the coupling of surface charge and ion flux asymmetries, and uncover the optimal surface configuration for enhanced self-propulsion. We find that nanoparticles with the sizes of $\sim 10-100$ nm in solutions with bulk ion concentrations of $\lesssim 5-50$ mM can show speeds that depend on surface configuration. For instance, taking the estimated values of $I \sim 10^7 $/s and $Q \sim - 10^2 \ e$ based on previous works \cite{DeCorato2020,Brown2017}, we find differences in speeds of up to $\sim 10$ $\mu$m/s or higher at bulk ion concentrations $<1$ mM for a particle of size $R=20$ nm (see Fig.~\ref{f3}). In particular, we find case (iii) produces the largest speed that becomes substantially larger with respect to the cases (i) and (ii) when $\kappa R \lesssim 1$. We obtain optimally high speeds in this case which can reach speeds $\gtrapprox 30$ $\mu$m/s at very low bulk concentrations $\lessapprox 0.01$ mM, while still maintaining speeds $\approx 1$ $\mu$m/s at much higher bulk concentrations $\approx 10$ mM. In contrast, we find very low speeds for case (iv) which shows an slight increase in speed as bulk concentrations is increased (see Fig.~\ref{f3}). Taken together, this indicates that the surface properties such as arrangement of surface charge and ion flux are crucial in regimes where the particle size is below or in the order of the Debye length. 

\begin{figure*}[t!]
	\center
	\includegraphics[width=1.6\columnwidth]{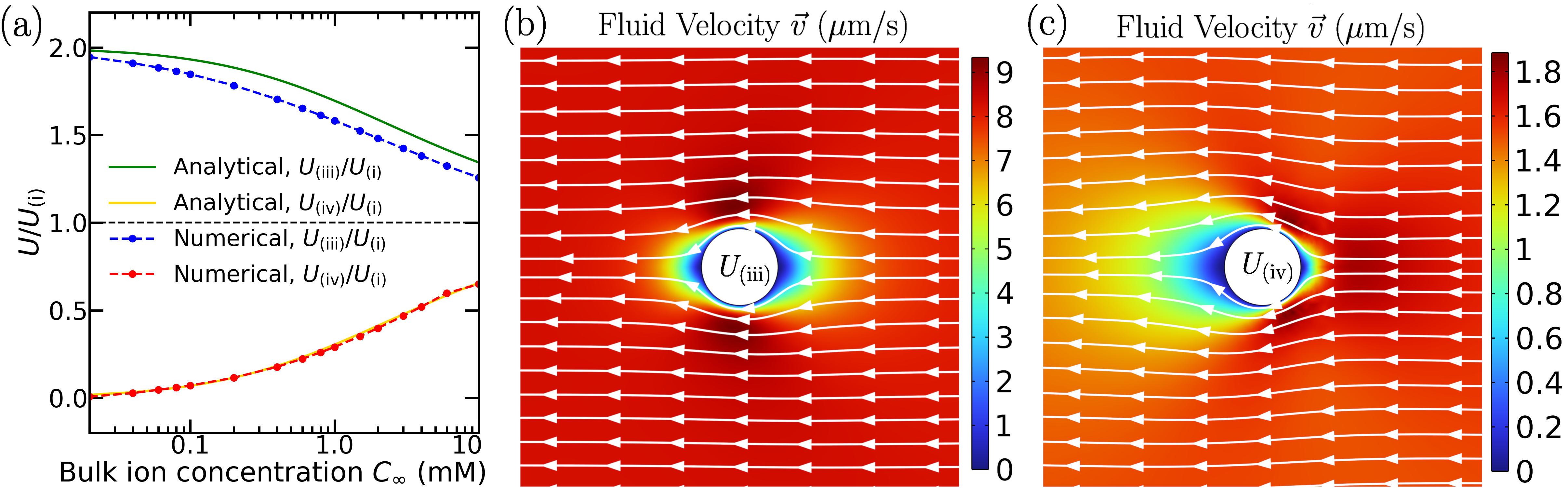} 
	\caption{Shifts in phoretic speed from coupling of surface charge and ion flux asymmetries. (a) The analytically estimated and numerically computed phoretic speed ratio with respect to case (i), $U/U_{\text{(i)}}$, as a function of bulk concentration $C_\infty$ for cases (iii) and (iv). The analytical estimates are obtained using Eq.~(\ref{eqE1}) and the numerical results are taken from Fig.~\ref{f3}. The color maps and field lines, from simulations in {\it COMSOL Multiphysics} \cite{comsol}, of the magnitude and flow lines of fluid velocity in the pariticle's rest frame for (b) case (iii) and (c) case (iv), both at $C_\infty=1$ nM. In all panels, we use $I \approx 6 \times 10^6$/s and $Q \approx - 60 e$ with $R=20$ nm, $D_+=1.98 \times 10^{-9}$ m$^2/$s, and $D_-=5.27 \times 10^{-9}$ m$^2/$s. 
	}
	\label{f4}
\end{figure*}

The shifts in the phoretic speeds arising from the surface charge and ionic flux coupling in case (iii) and (iv) can be realized through analytical estimations of the phoretic velocity ratios with respect to case (i) based on the leading order DH approximation, and examining the contrast in the induced fluid flow. In case (i), it has been shown in previous works  \cite{DeCorato2020,Asmolov2022,Brown2017} that the leading order phoretic velocity can be obtained using an ion flux $j(\theta)=j_0(1 + \cos{\theta})$ as it accounts for the leading monopolar and dipolar contributions of a Janus flux. Henceforth, we use this approximation to estimate the speed in case (i), and in a similar way, for a Janus surface charge in case (ii) we use Eq.~(\ref{eqUs}) as the speed. We find, when ignoring smaller cross terms, the speeds for case (iii) and (iv) can be estimated as linear combinations of the speeds for case (i) and (ii), and expressed as follows
\begin{equation} \label{eqE1}
    U_{\text{(iii)}} \approx (1 + \Delta(\kappa R)) U_{\text{(i)}}\,, \ \ \ \ \  U_{\text{(iv)}} \approx (1 - \Delta(\kappa R)) U_{\text{(i)}}  \,,
\end{equation}
where $\Delta(x)= 2(x+1)/(x^2+2x+2)$ such that $0 \leq \Delta (\kappa R) \leq 1$. These analytical predictions, that are consistent with the numerical results obtained from FE simulations as shown in Fig.~\ref{f4}(a), capture the amplifying and diminishing effects in cases (iii) and (iv) which lead to the respective shifts in speeds. Case (iii) (case (iv)) has aligning (opposing) leading velocity contributions that amplify (diminish) speed, and can be enhanced up to two folds compared to case (i) or (ii). Furthermore, figs.~\ref{f4}(b) and~\ref{f4}(c) show the resulting fluid flow in the particle's rest frame for case (iii) and (iv), respectively. Here, we find around the particle going from the moving front to the tail end, the fluid velocity increases in case (iv) and decreases in case (iv). This indicates the amplifying and diminishing effects emerging from the ion flux and surface charge interactions that directs the fluid around the particle inducing faster and slower self-propulsion in the respective cases.     

In this Letter, we find that surface charge asymmetry in chemically active colloids provides an alternative and additional mechanism through which a local ionic gradient is generated to induce self-propulsion. We show, in regimes where the colloid size becomes smaller than or comparable to Debye length, the near-field electrostatics and hydrodynamics play a crucial role in facilitating the self-phoretic motion that  arises uniquely from surface charge asymmetry, even when the ionic flux on the surface is uniform. The coupling of surface charge and ion flux asymmetries in Janus nanoparticles, with non-vanishing surface charge and ion flux placed on opposite hemispheres, leads to significantly enhanced phoretic speeds in the order of $\mu$m/s or higher. Taken together, our results suggest a new mode of phoretic self-propulsion in ionic media that can be utilized in the design of nanomotors, and moreover, lend insights to the underlying mechanism behind subcellular transport of nanoscale compartments with heterogeneous surface activity.  

This work has been supported by the Department of Energy (DOE), Office of Basic Energy Sciences under Contract DE-FG02-08ER46539. M.O.d.l.C. is thankful to the Sherman Fairchild Foundation for computational support.

\bibliographystyle{apsrev4-2}
\bibliography{ref}

\begin{thebibliography}{33}%
\makeatletter
\providecommand \@ifxundefined [1]{%
 \@ifx{#1\undefined}
}%
\providecommand \@ifnum [1]{%
 \ifnum #1\expandafter \@firstoftwo
 \else \expandafter \@secondoftwo
 \fi
}%
\providecommand \@ifx [1]{%
 \ifx #1\expandafter \@firstoftwo
 \else \expandafter \@secondoftwo
 \fi
}%
\providecommand \natexlab [1]{#1}%
\providecommand \enquote  [1]{``#1''}%
\providecommand \bibnamefont  [1]{#1}%
\providecommand \bibfnamefont [1]{#1}%
\providecommand \citenamefont [1]{#1}%
\providecommand \href@noop [0]{\@secondoftwo}%
\providecommand \href [0]{\begingroup \@sanitize@url \@href}%
\providecommand \@href[1]{\@@startlink{#1}\@@href}%
\providecommand \@@href[1]{\endgroup#1\@@endlink}%
\providecommand \@sanitize@url [0]{\catcode `\\12\catcode `\$12\catcode
  `\&12\catcode `\#12\catcode `\^12\catcode `\_12\catcode `\%12\relax}%
\providecommand \@@startlink[1]{}%
\providecommand \@@endlink[0]{}%
\providecommand \url  [0]{\begingroup\@sanitize@url \@url }%
\providecommand \@url [1]{\endgroup\@href {#1}{\urlprefix }}%
\providecommand \urlprefix  [0]{URL }%
\providecommand \Eprint [0]{\href }%
\providecommand \doibase [0]{https://doi.org/}%
\providecommand \selectlanguage [0]{\@gobble}%
\providecommand \bibinfo  [0]{\@secondoftwo}%
\providecommand \bibfield  [0]{\@secondoftwo}%
\providecommand \translation [1]{[#1]}%
\providecommand \BibitemOpen [0]{}%
\providecommand \bibitemStop [0]{}%
\providecommand \bibitemNoStop [0]{.\EOS\space}%
\providecommand \EOS [0]{\spacefactor3000\relax}%
\providecommand \BibitemShut  [1]{\csname bibitem#1\endcsname}%
\let\auto@bib@innerbib\@empty
\bibitem [{\citenamefont {Bray}(2000)}]{Bray2000}%
  \BibitemOpen
  \bibfield  {author} {\bibinfo {author} {\bibfnamefont {D.}~\bibnamefont
  {Bray}},\ }\href@noop {} {\emph {\bibinfo {title} {From Molecules to
  Motility}}},\ \bibinfo {edition} {2nd}\ ed.\ (\bibinfo  {publisher} {Garland
  Science},\ \bibinfo {year} {2000})\BibitemShut {NoStop}%
\bibitem [{\citenamefont {Lauga}\ and\ \citenamefont
  {Powers}(2009)}]{Lauga2009}%
  \BibitemOpen
  \bibfield  {author} {\bibinfo {author} {\bibfnamefont {E.}~\bibnamefont
  {Lauga}}\ and\ \bibinfo {author} {\bibfnamefont {T.~R.}\ \bibnamefont
  {Powers}},\ }\href@noop {} {\bibfield  {journal} {\bibinfo  {journal} {Rep.
  Prog. Phys.}\ }\textbf {\bibinfo {volume} {72}},\ \bibinfo {pages} {096601}
  (\bibinfo {year} {2009})}\BibitemShut {NoStop}%
\bibitem [{\citenamefont {Moran}\ and\ \citenamefont
  {Posner}(2017)}]{Moran2017}%
  \BibitemOpen
  \bibfield  {author} {\bibinfo {author} {\bibfnamefont {J.~L.}\ \bibnamefont
  {Moran}}\ and\ \bibinfo {author} {\bibfnamefont {J.~D.}\ \bibnamefont
  {Posner}},\ }\href@noop {} {\bibfield  {journal} {\bibinfo  {journal} {Annu.
  Rev. Fluid Mech.}\ }\textbf {\bibinfo {volume} {49}},\ \bibinfo {pages} {511}
  (\bibinfo {year} {2017})}\BibitemShut {NoStop}%
\bibitem [{\citenamefont {Gao}\ and\ \citenamefont {Wang}(2014)}]{Gao2014}%
  \BibitemOpen
  \bibfield  {author} {\bibinfo {author} {\bibfnamefont {W.}~\bibnamefont
  {Gao}}\ and\ \bibinfo {author} {\bibfnamefont {J.}~\bibnamefont {Wang}},\
  }\href@noop {} {\bibfield  {journal} {\bibinfo  {journal} {Nanoscale}\
  }\textbf {\bibinfo {volume} {6}},\ \bibinfo {pages} {10486} (\bibinfo {year}
  {2014})}\BibitemShut {NoStop}%
\bibitem [{\citenamefont {Soler}\ and\ \citenamefont
  {Sánchez}(2014)}]{Soler2014}%
  \BibitemOpen
  \bibfield  {author} {\bibinfo {author} {\bibfnamefont {L.}~\bibnamefont
  {Soler}}\ and\ \bibinfo {author} {\bibfnamefont {S.}~\bibnamefont
  {Sánchez}},\ }\href@noop {} {\bibfield  {journal} {\bibinfo  {journal}
  {Nanoscale}\ }\textbf {\bibinfo {volume} {6}},\ \bibinfo {pages} {7175}
  (\bibinfo {year} {2014})}\BibitemShut {NoStop}%
\bibitem [{\citenamefont {Anderson}(1989)}]{Anderson1989}%
  \BibitemOpen
  \bibfield  {author} {\bibinfo {author} {\bibfnamefont {J.~L.}\ \bibnamefont
  {Anderson}},\ }\href@noop {} {\bibfield  {journal} {\bibinfo  {journal}
  {Annu. Rev. Fluid Mech.}\ }\textbf {\bibinfo {volume} {21}},\ \bibinfo
  {pages} {61} (\bibinfo {year} {1989})}\BibitemShut {NoStop}%
\bibitem [{\citenamefont {Velegol}\ \emph {et~al.}(2016)\citenamefont
  {Velegol}, \citenamefont {Garg}, \citenamefont {Guha}, \citenamefont {Kar},\
  and\ \citenamefont {Kumar}}]{Velegol2016}%
  \BibitemOpen
  \bibfield  {author} {\bibinfo {author} {\bibfnamefont {D.}~\bibnamefont
  {Velegol}}, \bibinfo {author} {\bibfnamefont {A.}~\bibnamefont {Garg}},
  \bibinfo {author} {\bibfnamefont {R.}~\bibnamefont {Guha}}, \bibinfo {author}
  {\bibfnamefont {A.}~\bibnamefont {Kar}},\ and\ \bibinfo {author}
  {\bibfnamefont {M.}~\bibnamefont {Kumar}},\ }\href@noop {} {\bibfield
  {journal} {\bibinfo  {journal} {Soft Matter}\ }\textbf {\bibinfo {volume}
  {12}},\ \bibinfo {pages} {4686} (\bibinfo {year} {2016})}\BibitemShut
  {NoStop}%
\bibitem [{\citenamefont {Marchetti}\ \emph {et~al.}(2013)\citenamefont
  {Marchetti}, \citenamefont {Joanny}, \citenamefont {Ramaswamy}, \citenamefont
  {Liverpool}, \citenamefont {Prost}, \citenamefont {Rao},\ and\ \citenamefont
  {Simha}}]{Marchetti2013}%
  \BibitemOpen
  \bibfield  {author} {\bibinfo {author} {\bibfnamefont {M.~C.}\ \bibnamefont
  {Marchetti}}, \bibinfo {author} {\bibfnamefont {J.~F.}\ \bibnamefont
  {Joanny}}, \bibinfo {author} {\bibfnamefont {S.}~\bibnamefont {Ramaswamy}},
  \bibinfo {author} {\bibfnamefont {T.~B.}\ \bibnamefont {Liverpool}}, \bibinfo
  {author} {\bibfnamefont {J.}~\bibnamefont {Prost}}, \bibinfo {author}
  {\bibfnamefont {M.}~\bibnamefont {Rao}},\ and\ \bibinfo {author}
  {\bibfnamefont {R.~A.}\ \bibnamefont {Simha}},\ }\href@noop {} {\bibfield
  {journal} {\bibinfo  {journal} {Rev. Mod. Phys.}\ }\textbf {\bibinfo {volume}
  {85}},\ \bibinfo {pages} {1143} (\bibinfo {year} {2013})}\BibitemShut
  {NoStop}%
\bibitem [{\citenamefont {Illien}\ \emph {et~al.}(2017)\citenamefont {Illien},
  \citenamefont {Golestanian},\ and\ \citenamefont {Sen}}]{Illien2017}%
  \BibitemOpen
  \bibfield  {author} {\bibinfo {author} {\bibfnamefont {P.}~\bibnamefont
  {Illien}}, \bibinfo {author} {\bibfnamefont {R.}~\bibnamefont
  {Golestanian}},\ and\ \bibinfo {author} {\bibfnamefont {A.}~\bibnamefont
  {Sen}},\ }\href@noop {} {\bibfield  {journal} {\bibinfo  {journal} {Chem.
  Soc. Rev.}\ }\textbf {\bibinfo {volume} {46}},\ \bibinfo {pages} {5508}
  (\bibinfo {year} {2017})}\BibitemShut {NoStop}%
\bibitem [{\citenamefont {Golestanian}\ \emph {et~al.}(2005)\citenamefont
  {Golestanian}, \citenamefont {Liverpool},\ and\ \citenamefont
  {Ajdari}}]{Golestanian2005}%
  \BibitemOpen
  \bibfield  {author} {\bibinfo {author} {\bibfnamefont {R.}~\bibnamefont
  {Golestanian}}, \bibinfo {author} {\bibfnamefont {T.~B.}\ \bibnamefont
  {Liverpool}},\ and\ \bibinfo {author} {\bibfnamefont {A.}~\bibnamefont
  {Ajdari}},\ }\href@noop {} {\bibfield  {journal} {\bibinfo  {journal} {Phys.
  Rev. Lett.}\ }\textbf {\bibinfo {volume} {94}},\ \bibinfo {pages} {220801}
  (\bibinfo {year} {2005})}\BibitemShut {NoStop}%
\bibitem [{\citenamefont {Golestanian}\ \emph {et~al.}(2007)\citenamefont
  {Golestanian}, \citenamefont {Liverpool},\ and\ \citenamefont
  {Ajdari}}]{Golestanian2007}%
  \BibitemOpen
  \bibfield  {author} {\bibinfo {author} {\bibfnamefont {R.}~\bibnamefont
  {Golestanian}}, \bibinfo {author} {\bibfnamefont {T.~B.}\ \bibnamefont
  {Liverpool}},\ and\ \bibinfo {author} {\bibfnamefont {A.}~\bibnamefont
  {Ajdari}},\ }\href@noop {} {\bibfield  {journal} {\bibinfo  {journal} {New J.
  Phys.}\ }\textbf {\bibinfo {volume} {9}},\ \bibinfo {pages} {126} (\bibinfo
  {year} {2007})}\BibitemShut {NoStop}%
\bibitem [{\citenamefont {Zhang}\ \emph {et~al.}(2017)\citenamefont {Zhang},
  \citenamefont {Grzybowski},\ and\ \citenamefont {Granick}}]{Zhang2017}%
  \BibitemOpen
  \bibfield  {author} {\bibinfo {author} {\bibfnamefont {J.}~\bibnamefont
  {Zhang}}, \bibinfo {author} {\bibfnamefont {B.~A.}\ \bibnamefont
  {Grzybowski}},\ and\ \bibinfo {author} {\bibfnamefont {S.}~\bibnamefont
  {Granick}},\ }\href@noop {} {\bibfield  {journal} {\bibinfo  {journal}
  {Langmuir}\ }\textbf {\bibinfo {volume} {33}},\ \bibinfo {pages} {6964}
  (\bibinfo {year} {2017})}\BibitemShut {NoStop}%
\bibitem [{\citenamefont {Zhou}\ \emph {et~al.}(2018)\citenamefont {Zhou},
  \citenamefont {Zhang}, \citenamefont {Tang},\ and\ \citenamefont
  {Wang}}]{Zhou2018}%
  \BibitemOpen
  \bibfield  {author} {\bibinfo {author} {\bibfnamefont {C.}~\bibnamefont
  {Zhou}}, \bibinfo {author} {\bibfnamefont {H.~P.}\ \bibnamefont {Zhang}},
  \bibinfo {author} {\bibfnamefont {J.}~\bibnamefont {Tang}},\ and\ \bibinfo
  {author} {\bibfnamefont {W.}~\bibnamefont {Wang}},\ }\href@noop {} {\bibfield
   {journal} {\bibinfo  {journal} {Langmuir}\ }\textbf {\bibinfo {volume}
  {34}},\ \bibinfo {pages} {3289} (\bibinfo {year} {2018})}\BibitemShut
  {NoStop}%
\bibitem [{\citenamefont {De~Corato}\ \emph {et~al.}(2020)\citenamefont
  {De~Corato}, \citenamefont {Arqu\'e}, \citenamefont {Pati\~no}, \citenamefont
  {Arroyo}, \citenamefont {S\'anchez},\ and\ \citenamefont
  {Pagonabarraga}}]{DeCorato2020}%
  \BibitemOpen
  \bibfield  {author} {\bibinfo {author} {\bibfnamefont {M.}~\bibnamefont
  {De~Corato}}, \bibinfo {author} {\bibfnamefont {X.}~\bibnamefont {Arqu\'e}},
  \bibinfo {author} {\bibfnamefont {T.}~\bibnamefont {Pati\~no}}, \bibinfo
  {author} {\bibfnamefont {M.}~\bibnamefont {Arroyo}}, \bibinfo {author}
  {\bibfnamefont {S.}~\bibnamefont {S\'anchez}},\ and\ \bibinfo {author}
  {\bibfnamefont {I.}~\bibnamefont {Pagonabarraga}},\ }\href@noop {} {\bibfield
   {journal} {\bibinfo  {journal} {Phys. Rev. Lett.}\ }\textbf {\bibinfo
  {volume} {124}},\ \bibinfo {pages} {108001} (\bibinfo {year}
  {2020})}\BibitemShut {NoStop}%
\bibitem [{\citenamefont {Asmolov}\ \emph {et~al.}(2022)\citenamefont
  {Asmolov}, \citenamefont {Nizkaya},\ and\ \citenamefont
  {Vinogradova}}]{Asmolov2022}%
  \BibitemOpen
  \bibfield  {author} {\bibinfo {author} {\bibfnamefont {E.~S.}\ \bibnamefont
  {Asmolov}}, \bibinfo {author} {\bibfnamefont {T.~V.}\ \bibnamefont
  {Nizkaya}},\ and\ \bibinfo {author} {\bibfnamefont {O.~I.}\ \bibnamefont
  {Vinogradova}},\ }\href@noop {} {\bibfield  {journal} {\bibinfo  {journal}
  {Phys. Fluids}\ }\textbf {\bibinfo {volume} {34}},\ \bibinfo {pages} {032011}
  (\bibinfo {year} {2022})}\BibitemShut {NoStop}%
\bibitem [{\citenamefont {Brown}\ \emph {et~al.}(2017)\citenamefont {Brown},
  \citenamefont {Poon}, \citenamefont {Holm},\ and\ \citenamefont
  {de~Graaf}}]{Brown2017}%
  \BibitemOpen
  \bibfield  {author} {\bibinfo {author} {\bibfnamefont {A.~T.}\ \bibnamefont
  {Brown}}, \bibinfo {author} {\bibfnamefont {W.~C.~K.}\ \bibnamefont {Poon}},
  \bibinfo {author} {\bibfnamefont {C.}~\bibnamefont {Holm}},\ and\ \bibinfo
  {author} {\bibfnamefont {J.}~\bibnamefont {de~Graaf}},\ }\href@noop {}
  {\bibfield  {journal} {\bibinfo  {journal} {Soft Matter}\ }\textbf {\bibinfo
  {volume} {13}},\ \bibinfo {pages} {1200} (\bibinfo {year}
  {2017})}\BibitemShut {NoStop}%
\bibitem [{\citenamefont {Hermanov\'{a}}\ and\ \citenamefont
  {Pumera}(2020)}]{Hermanova2020}%
  \BibitemOpen
  \bibfield  {author} {\bibinfo {author} {\bibfnamefont {S.}~\bibnamefont
  {Hermanov\'{a}}}\ and\ \bibinfo {author} {\bibfnamefont {M.}~\bibnamefont
  {Pumera}},\ }\href@noop {} {\bibfield  {journal} {\bibinfo  {journal} {Chem.
  Euro. J.}\ }\textbf {\bibinfo {volume} {26}},\ \bibinfo {pages} {11085}
  (\bibinfo {year} {2020})}\BibitemShut {NoStop}%
\bibitem [{\citenamefont {Safdar}\ \emph {et~al.}(2018)\citenamefont {Safdar},
  \citenamefont {Khan},\ and\ \citenamefont {Jänis}}]{Safdar2018}%
  \BibitemOpen
  \bibfield  {author} {\bibinfo {author} {\bibfnamefont {M.}~\bibnamefont
  {Safdar}}, \bibinfo {author} {\bibfnamefont {S.~U.}\ \bibnamefont {Khan}},\
  and\ \bibinfo {author} {\bibfnamefont {J.}~\bibnamefont {Jänis}},\
  }\href@noop {} {\bibfield  {journal} {\bibinfo  {journal} {Adv. Mater.}\
  }\textbf {\bibinfo {volume} {30}},\ \bibinfo {pages} {1703660} (\bibinfo
  {year} {2018})}\BibitemShut {NoStop}%
\bibitem [{\citenamefont {Moran}\ and\ \citenamefont
  {Posner}(2011)}]{Moran2011}%
  \BibitemOpen
  \bibfield  {author} {\bibinfo {author} {\bibfnamefont {J.~L.}\ \bibnamefont
  {Moran}}\ and\ \bibinfo {author} {\bibfnamefont {J.~D.}\ \bibnamefont
  {Posner}},\ }\href@noop {} {\bibfield  {journal} {\bibinfo  {journal} {J.
  Fluid Mech.}\ }\textbf {\bibinfo {volume} {680}},\ \bibinfo {pages} {31–66}
  (\bibinfo {year} {2011})}\BibitemShut {NoStop}%
\bibitem [{\citenamefont {Sutter}\ \emph {et~al.}(2021)\citenamefont {Sutter},
  \citenamefont {Melnicki}, \citenamefont {Schulz}, \citenamefont {Woyke},\
  and\ \citenamefont {Kerfeld}}]{Sutter2021}%
  \BibitemOpen
  \bibfield  {author} {\bibinfo {author} {\bibfnamefont {M.}~\bibnamefont
  {Sutter}}, \bibinfo {author} {\bibfnamefont {M.~R.}\ \bibnamefont
  {Melnicki}}, \bibinfo {author} {\bibfnamefont {F.}~\bibnamefont {Schulz}},
  \bibinfo {author} {\bibfnamefont {T.}~\bibnamefont {Woyke}},\ and\ \bibinfo
  {author} {\bibfnamefont {C.~A.}\ \bibnamefont {Kerfeld}},\ }\href@noop {}
  {\bibfield  {journal} {\bibinfo  {journal} {Nat. Commun.}\ }\textbf {\bibinfo
  {volume} {12}},\ \bibinfo {pages} {3809} (\bibinfo {year}
  {2021})}\BibitemShut {NoStop}%
\bibitem [{\citenamefont {Kerfeld}\ \emph {et~al.}(2005)\citenamefont
  {Kerfeld}, \citenamefont {Sawaya}, \citenamefont {Tanaka}, \citenamefont
  {Nguyen}, \citenamefont {Phillips}, \citenamefont {Beeby},\ and\
  \citenamefont {Yeates}}]{Kerfeld2005}%
  \BibitemOpen
  \bibfield  {author} {\bibinfo {author} {\bibfnamefont {C.~A.}\ \bibnamefont
  {Kerfeld}}, \bibinfo {author} {\bibfnamefont {M.~R.}\ \bibnamefont {Sawaya}},
  \bibinfo {author} {\bibfnamefont {S.}~\bibnamefont {Tanaka}}, \bibinfo
  {author} {\bibfnamefont {C.~V.}\ \bibnamefont {Nguyen}}, \bibinfo {author}
  {\bibfnamefont {M.}~\bibnamefont {Phillips}}, \bibinfo {author}
  {\bibfnamefont {M.}~\bibnamefont {Beeby}},\ and\ \bibinfo {author}
  {\bibfnamefont {T.~O.}\ \bibnamefont {Yeates}},\ }\href@noop {} {\bibfield
  {journal} {\bibinfo  {journal} {Science}\ }\textbf {\bibinfo {volume}
  {309}},\ \bibinfo {pages} {936} (\bibinfo {year} {2005})}\BibitemShut
  {NoStop}%
\bibitem [{\citenamefont {Sommer}\ \emph {et~al.}(2017)\citenamefont {Sommer},
  \citenamefont {Cai}, \citenamefont {Melnicki},\ and\ \citenamefont
  {Kerfeld}}]{Sommer2017}%
  \BibitemOpen
  \bibfield  {author} {\bibinfo {author} {\bibfnamefont {M.}~\bibnamefont
  {Sommer}}, \bibinfo {author} {\bibfnamefont {F.}~\bibnamefont {Cai}},
  \bibinfo {author} {\bibfnamefont {M.}~\bibnamefont {Melnicki}},\ and\
  \bibinfo {author} {\bibfnamefont {C.~A.}\ \bibnamefont {Kerfeld}},\
  }\href@noop {} {\bibfield  {journal} {\bibinfo  {journal} {J. Exp. Bot.}\
  }\textbf {\bibinfo {volume} {68}},\ \bibinfo {pages} {3841} (\bibinfo {year}
  {2017})}\BibitemShut {NoStop}%
\bibitem [{\citenamefont {Sutter}\ \emph {et~al.}(2019)\citenamefont {Sutter},
  \citenamefont {Laughlin}, \citenamefont {Sloan}, \citenamefont {Serwas},
  \citenamefont {Davies},\ and\ \citenamefont {Kerfeld}}]{Sutter2019}%
  \BibitemOpen
  \bibfield  {author} {\bibinfo {author} {\bibfnamefont {M.}~\bibnamefont
  {Sutter}}, \bibinfo {author} {\bibfnamefont {T.~G.}\ \bibnamefont
  {Laughlin}}, \bibinfo {author} {\bibfnamefont {N.~B.}\ \bibnamefont {Sloan}},
  \bibinfo {author} {\bibfnamefont {D.}~\bibnamefont {Serwas}}, \bibinfo
  {author} {\bibfnamefont {K.~M.}\ \bibnamefont {Davies}},\ and\ \bibinfo
  {author} {\bibfnamefont {C.~A.}\ \bibnamefont {Kerfeld}},\ }\href@noop {}
  {\bibfield  {journal} {\bibinfo  {journal} {Plant Physiol.}\ }\textbf
  {\bibinfo {volume} {181}},\ \bibinfo {pages} {1050} (\bibinfo {year}
  {2019})}\BibitemShut {NoStop}%
\bibitem [{\citenamefont {Santiago}(2018)}]{Santiago2018}%
  \BibitemOpen
  \bibfield  {author} {\bibinfo {author} {\bibfnamefont {I.}~\bibnamefont
  {Santiago}},\ }\href@noop {} {\bibfield  {journal} {\bibinfo  {journal} {Nano
  Today}\ }\textbf {\bibinfo {volume} {19}},\ \bibinfo {pages} {11} (\bibinfo
  {year} {2018})}\BibitemShut {NoStop}%
\bibitem [{\citenamefont {Lee}\ \emph {et~al.}(2014)\citenamefont {Lee},
  \citenamefont {Alarcón-Correa}, \citenamefont {Miksch}, \citenamefont
  {Hahn}, \citenamefont {Gibbs},\ and\ \citenamefont {Fischer}}]{Lee2014}%
  \BibitemOpen
  \bibfield  {author} {\bibinfo {author} {\bibfnamefont {T.-C.}\ \bibnamefont
  {Lee}}, \bibinfo {author} {\bibfnamefont {M.}~\bibnamefont
  {Alarcón-Correa}}, \bibinfo {author} {\bibfnamefont {C.}~\bibnamefont
  {Miksch}}, \bibinfo {author} {\bibfnamefont {K.}~\bibnamefont {Hahn}},
  \bibinfo {author} {\bibfnamefont {J.~G.}\ \bibnamefont {Gibbs}},\ and\
  \bibinfo {author} {\bibfnamefont {P.}~\bibnamefont {Fischer}},\ }\href@noop
  {} {\bibfield  {journal} {\bibinfo  {journal} {Nano Lett.}\ }\textbf
  {\bibinfo {volume} {14}},\ \bibinfo {pages} {2407} (\bibinfo {year}
  {2014})}\BibitemShut {NoStop}%
\bibitem [{\citenamefont {Ramm}\ \emph {et~al.}(2021)\citenamefont {Ramm},
  \citenamefont {Goychuk}, \citenamefont {Khmelinskaia}, \citenamefont
  {Blumhardt}, \citenamefont {Eto}, \citenamefont {Ganzinger}, \citenamefont
  {Frey},\ and\ \citenamefont {Schwille}}]{Ramm2021}%
  \BibitemOpen
  \bibfield  {author} {\bibinfo {author} {\bibfnamefont {B.}~\bibnamefont
  {Ramm}}, \bibinfo {author} {\bibfnamefont {A.}~\bibnamefont {Goychuk}},
  \bibinfo {author} {\bibfnamefont {A.}~\bibnamefont {Khmelinskaia}}, \bibinfo
  {author} {\bibfnamefont {P.}~\bibnamefont {Blumhardt}}, \bibinfo {author}
  {\bibfnamefont {H.}~\bibnamefont {Eto}}, \bibinfo {author} {\bibfnamefont
  {K.~A.}\ \bibnamefont {Ganzinger}}, \bibinfo {author} {\bibfnamefont
  {E.}~\bibnamefont {Frey}},\ and\ \bibinfo {author} {\bibfnamefont
  {P.}~\bibnamefont {Schwille}},\ }\href@noop {} {\bibfield  {journal}
  {\bibinfo  {journal} {Nat. Phys.}\ }\textbf {\bibinfo {volume} {17}},\
  \bibinfo {pages} {850} (\bibinfo {year} {2021})}\BibitemShut {NoStop}%
\bibitem [{\citenamefont {Sear}(2019)}]{Sear2019}%
  \BibitemOpen
  \bibfield  {author} {\bibinfo {author} {\bibfnamefont {R.~P.}\ \bibnamefont
  {Sear}},\ }\href@noop {} {\bibfield  {journal} {\bibinfo  {journal} {Phys.
  Rev. Lett.}\ }\textbf {\bibinfo {volume} {122}},\ \bibinfo {pages} {128101}
  (\bibinfo {year} {2019})}\BibitemShut {NoStop}%
\bibitem [{\citenamefont {Yariv}(2011)}]{Yariv2011}%
  \BibitemOpen
  \bibfield  {author} {\bibinfo {author} {\bibfnamefont {E.}~\bibnamefont
  {Yariv}},\ }\href@noop {} {\bibfield  {journal} {\bibinfo  {journal} {Proc.
  R. Soc. A}\ }\textbf {\bibinfo {volume} {467}},\ \bibinfo {pages} {1645}
  (\bibinfo {year} {2011})}\BibitemShut {NoStop}%
\bibitem [{\citenamefont {Happel}\ and\ \citenamefont
  {Brenner}(1983)}]{Happel1983}%
  \BibitemOpen
  \bibfield  {author} {\bibinfo {author} {\bibfnamefont {J.}~\bibnamefont
  {Happel}}\ and\ \bibinfo {author} {\bibfnamefont {H.}~\bibnamefont
  {Brenner}},\ }\href@noop {} {\emph {\bibinfo {title} {Low Reynolds number
  hydrodynamics}}}\ (\bibinfo  {publisher} {Springer Netherlands},\ \bibinfo
  {year} {1983})\BibitemShut {NoStop}%
\bibitem [{\citenamefont {Stone}\ and\ \citenamefont
  {Samuel}(1996)}]{Stone1996}%
  \BibitemOpen
  \bibfield  {author} {\bibinfo {author} {\bibfnamefont {H.~A.}\ \bibnamefont
  {Stone}}\ and\ \bibinfo {author} {\bibfnamefont {A.~D.~T.}\ \bibnamefont
  {Samuel}},\ }\href@noop {} {\bibfield  {journal} {\bibinfo  {journal} {Phys.
  Rev. Lett.}\ }\textbf {\bibinfo {volume} {77}},\ \bibinfo {pages} {4102}
  (\bibinfo {year} {1996})}\BibitemShut {NoStop}%
\bibitem [{\citenamefont {Masoud}\ and\ \citenamefont
  {Stone}(2019)}]{Masoud2019}%
  \BibitemOpen
  \bibfield  {author} {\bibinfo {author} {\bibfnamefont {H.}~\bibnamefont
  {Masoud}}\ and\ \bibinfo {author} {\bibfnamefont {H.~A.}\ \bibnamefont
  {Stone}},\ }\href@noop {} {\bibfield  {journal} {\bibinfo  {journal} {J.
  Fluid Mech.}\ }\textbf {\bibinfo {volume} {879}},\ \bibinfo {pages} {P1}
  (\bibinfo {year} {2019})}\BibitemShut {NoStop}%
\bibitem [{\citenamefont {Landau}\ and\ \citenamefont
  {Lifshitz}(1987)}]{Landau1987}%
  \BibitemOpen
  \bibfield  {author} {\bibinfo {author} {\bibfnamefont {L.}~\bibnamefont
  {Landau}}\ and\ \bibinfo {author} {\bibfnamefont {E.}~\bibnamefont
  {Lifshitz}},\ }\href@noop {} {\emph {\bibinfo {title} {Fluid Mechanics}}},\
  \bibinfo {edition} {2nd}\ ed.\ (\bibinfo  {publisher} {Pergamon},\ \bibinfo
  {year} {1987})\BibitemShut {NoStop}%
\bibitem [{com()}]{comsol}%
  \BibitemOpen
  \href@noop {} {\bibinfo  {journal} {COMSOL Multiphysics v. 6.0. COMSOL AB,
  Stockholm, Sweden. www.comsol.com}\ }\BibitemShut {NoStop}%
\end{thebibliography}%

\end{document}